\documentclass[letterpaper]{article} 
\usepackage{aaai24}  
\usepackage{times}  
\usepackage{helvet}  
\usepackage{courier}  
\usepackage[hyphens]{url}  
\usepackage{graphicx} 
\usepackage{natbib}  
\usepackage{caption} 
\usepackage{algorithm}
\usepackage{algorithmic}

\usepackage{newfloat}
\usepackage{listings}
\DeclareCaptionStyle{ruled}{labelfont=normalfont,labelsep=colon,strut=off} 
\floatstyle{ruled}
\newfloat{listing}{tb}{lst}{}
\floatname{listing}{Listing}
\title{A Toolbox for Modelling Engagement with Educational Videos}

\author{
Yuxiang Qiu\equalcontrib,
Karim Djemili\equalcontrib,
Denis Elezi\equalcontrib,
Aaneel Shalman Srazali\equalcontrib,
Mar\'ia P\'erez-Ortiz, 
Emine Yilmaz, 
John Shawe-Taylor and
Sahan Bulathwela
}

\affiliations{
Department of Computer Science, University College London \\
Gower Street, London WC1E 6BT, UK\\
m.bulathwela@ucl.ac.uk
}

\usepackage{bibentry}

\begin{document}

\maketitle

\begin{abstract}
With the advancement and utility of Artificial Intelligence (AI), personalising education to a global population could be a cornerstone of new educational systems in the future. This work presents the \emph{PEEKC} dataset and the \emph{TrueLearn} Python library, which contains a dataset and a series of online learner state models that are essential to facilitate research on learner engagement modelling. 
TrueLearn family of models was designed following the "open learner" concept, using humanly-intuitive user representations. 
This family of scalable, online models also help  
end-users visualise the learner models, which may in the future facilitate user interaction with their models/recommenders.
The extensive documentation and coding examples make the library highly accessible to both machine learning developers and educational data mining and learning analytics practitioners.  
The experiments show the utility of both the dataset and the library with predictive performance significantly exceeding comparative baseline models. The dataset contains a large amount of AI-related educational videos, which are of interest for building and validating AI-specific educational recommenders.
\end{abstract}

\section{Introduction}

It has been shown that personalised one-on-one learning could lead to improving learning gains by two standard deviations \cite{Bloom84}. 
With this goal in sight, and the ambition to democratise education to a world population, 
we require responsible intelligent systems that can bring scalable, personalised and governable models to a mass of learners \cite{democratise2021}.
Intelligent Tutoring Systems (ITS), the go-to solution, is practical for courses with a limited number of learning materials and heavily relies on testing users for knowledge. However, today's world has access to 100,000s of rich educational videos, PDFs and podcasts that can be matched to a global population of lifelong learners. 
Educational recommenders have the opportunity to 
leverage implicit interaction signals (such as clicks and watch time) to personalise and support learning for informal lifelong learners \cite{bulathwela2022sus}. Furthermore, {the scarcity of publicly available datasets of learners in the wild engaging with educational materials} is a major deterrent to creating scalable educational recommenders.

The contributions of this paper are two-fold. Firstly, we create and release to the public the \textbf{P}ersonalised \textbf{E}ducational \textbf{E}ngagement linked to \textbf{K}nowledge \textbf{C}omponents (PEEKC) dataset, with more than 20,000 informal learners watching educational videos in an in-the-wild setting (i.e. learning informally). The videos in the PEEKC dataset majorly contain concepts related to Artificial Intelligence (AI) and Machine Learning (ML) making it a valuable resource for AI-assisted education on these topics. Secondly, we develop and release \emph{TrueLearn},
an open-source Python library packaging state-of-the-art Bayesian models and visualisation tools for leveraging scalable, online learning, transparent learner models. The library contains different components that will enable i) creating content representations of learning resources ii) managing user/learner states, iii) modelling the state evolution of learners using interactions and iv) evaluating engagement predictions. This work uses the PEEKC dataset to empirically demonstrate the predictive capabilities of the different models within the TrueLearn library.

\section{Related Work}

The scarcity of publicly available datasets 
for predicting learner engagement with educational videos constrains the growth of the personalisation of AI education. PEEKC is the first and largest learner video engagement dataset publicly released with humanly interpretable Wikipedia concepts and the concept coverage associated with the video lecture fragments. 
Next, we present relevant works to i) the PEEKC dataset (not only related datasets but also research on the approaches that were used to generate it, e.g. Wikification) and ii) our novel machine learning library. 

\subsection{Related Datasets}

\emph{Knowledge Tracing} \cite{corbett1994knowledge,deep_kt}, focussing on modelling knowledge/skill mastery of learners based on test-taking is the most active research area in the learner modelling domain. ASSISTments data \cite{assistments_data}, which records learners solving mathematics problems, is often used in literature while this data is mathematics education focused. 
Additionally, 
problem-solving interactions \cite{choi2020ednet}, multiple choice question answering \cite{wang2020diagnostic,wang2021educational} datasets exist publicly while none of them includes implicit feedback related to consuming educational videos. 
MOOCCube dataset contains a spectrum of different statistics relating to learner-MOOC interactions including implicit and explicit test-taking activity \cite{yu2020mooccube}. Although this dataset may contain data that can be used to predict learner engagement which has been used for course recommendation \cite{deng2023knowledge}, the pre-organisation of courses takes away the in-the-wild choices learners make in choosing videos/fragments to watch while the courses in MOOCCube are not limited to AI. On the contrary, PEEKC dataset presents over 20,000 informal learners watching AI-related videos with fragment-level annotation of videos providing more granularity at a time segment/fragment level information retrieval is gaining interest \cite{yu2020spotify}. 

\subsection{Extracting Knowledge Components}

ITSs often rely on expert labelling of the \textbf{Knowledge Components (KCs)} \cite{assistments_data} 
, which is time-consuming and not scalable. Unsupervised learning approaches
are also potential candidates.
{Latent Dirichlet Allocation (LDA) has widely been used to extract topic metadata from different types of text data including course syllabuses \cite{Apaza2014OnlineCR}. However, unsupervised approaches such as LDA suffer from complex hyper-parameter tuning \cite{PANICHELLA2021106411}} and limited interpretability of \emph{latent} KCs, creating gaps in transparency. 
\emph{Wikification}, a form of entity linking \cite{wikifier} 
has shown promise for automatically capturing the KCs covered in an educational resource \cite{truelearn}. This technology provides \emph{automatic}, \emph{humanly-intuitive (symbolic)} representations from Wikipedia, representing \emph{up-to-date knowledge} about \emph{many domains}. 

The possibility of recommending parts of items (contrary to an entire video or podcast) is also a fruitful research direction explored lately. From proximity-aware information retrieval 
\cite{schenkel2007efficient} to segmenting videos to build tables of contents \cite{videoken}, this goal has been under active research. Breaking informational videos into fragments has also shown promise in efficient previewing \cite{chen2018temporally} and enabling non-linear consumption of videos \cite{non_lin}. Recent proposals such as TrueLearn \cite{truelearn} demonstrate the potential of using fragment recommendation in education. 
Due to these reasons, we use Wikification \cite{wikifier} to generate KCs that are included in each video fragment covered by the PEEKC dataset.

\subsection{Designing a Machine Learning Library}

To design a user-friendly, easy-to-use, and scalable library, commonly used yet, bad design practices, such as rigidity, fragility, immobility and viscosity should be avoided \cite{martin2000design,piccioni2013empirical}. 
Many design principles are proposed to overcome these issues \cite{gamma1995design}. Many data scientists also prefer usable (adhering to known patterns), well-documented and intuitive libraries \cite{nadi2023selecting}.
Besides these, designing a machine learning library entails overcoming additional challenges (e.g. data, pre-processing, models, etc.).
Scikit-learn \cite{pedregosa2011scikit} proposes consistency, inspection, sensible defaults and good interface design (estimators and predictors) for building a scalable and user-friendly machine learning library. Consistency of the code interfaces significantly reduces the learning cost for users while inspection exposes relevant model parameters and public attributes to the user with easy access \cite{buitinck2013api}.
The estimator interface specifies a \verb|fit| function to provide a consistent interface to the training model and exposes the \verb|coef_| attribute to facilitate the inspection of the internal state of the model. The predictor interface specifies the \verb|predict| and \verb|predict_proba| functions as methods for utilising the trained model. PyBKT, a Python-based library that implements knowledge tracing and item response theory-based learner models, also follow the same interfacing practices where function names \verb|fit| and \verb|evaluate| are used to train and predict \cite{psych5030050}. 
Due to the time-tested and consistent design decisions that have succeeded in scikit-learn and pyBKT, we utilise the same functions to interact with the learner models in the TrueLearn library.


\subsection{Learner Modelling}

Personalised learning mainly revolves around Knowledge Tracing (KT) \cite{psych5030050} and Item-Response Theory (IRT) \cite{Rasch1960} based models that use KCs in exercises to predict test success. However, these models focus on test-taking (modelling short sequences of exercise answering events) rather than consuming learning materials such as video watching. Conventional KT and IRT models do not support online learning posing scale challenges in lifelong learning cases (while online counterparts exist \cite{bishopsnewbook}). More recently, deep-KT \cite{deep_kt} has shown promise in superior performance. However, deep-KT models are data-hungry and lack interpretability, making them less favourable for lifelong learning, where the model needs to learn usable parameters with minimal data. Furthermore, recent studies have questioned the superior performance of deep-KT models in comparison to traditional models \cite{Schmucker_Wang_Hu_Mitchell_2022}. Due to these reasons, we scope out batch/deep learning models and focus on data-efficient online models.

The TrueLearn family of online Bayesian learner models uses implicit feedback from learners to recover their learning state \cite{truelearn}. Models that capture learners' interests, knowledge, and novelty are proposed in prior work with methods to combine them as interpretable ensembles that can account for these factors simultaneously \cite{bulathwela2022sus}. 
While being data efficient and privacy-preserving by design 
(exclusively using individual learner's interactions)
, 
TrueLearn models generate humanly intuitive learner representations inspired by Open Learner Models (OLM).
This involves generating visualisations that will communicate information about learner state, promoting learner reflection by aiding learners in planning and monitoring their learning \cite{openlearnermodels}. 
OLMs also pose challenges, since all visual presentations may not be equally understood by a wide variety of end-users. Among many visualisations used to present learner knowledge state, user studies have shown that some visualisations are comparatively more user-friendly than others \cite{10.1007/978-3-319-98572-5_40,visualisationscomparison}. TrueLearn implements a set of tested visualisations that aid the communication and the interaction process. 

Both deep KT and libraries such as pyBKT focus on predicting test-taking behaviour \cite{psych5030050} rather than how they would interact with an educational video. These libraries also focus on course-based learning settings where the number of KCs and learning items are limited in number. In these aspects, TrueLearn sets itself apart from the rest of the available libraries. The same reasons make TrueLearn valuable for MOOC platforms and educational video repositories that thrive to personalise videos for learning. In a world where a large number of educational videos are in circulation, we are unaware of a public, easy-to-use toolkit that can be used to incorporate educational video personalisation apart from our proposal, TrueLearn.


\section{Problem Setting}

A learner $\ell$ in learner population $L$ interacting with a series of educational resources $S_\ell \subset \{r_1, \ldots, r_{R}\}$ where $r_x$ are \emph{fragments/parts} of an educational video $v$. The watch interactions happen over a period of $T$ time steps, $R$ being the total number of resources in the system.
In this system with a total $N$ unique knowledge components (KCs), resource $r_x$ is characterised by a set of top KCs or topics $K_{r_x} \subset \{1, \ldots, N \}$. We assume the presence $i_{r_x}$ of KC in resource $r_x$ and the degree $d_{r_x}\in \{ 0,1\}$ of KC coverage in the resource is observable.

The key idea is to model the probability of engagement $e_{\ell, r_x}^{t} \in \{ 1, -1\}$ between learner $\ell$ and resource $r_x$ at time $t$ as a function of the learner interest $\theta^t_{\ell_{\texttt{I}}}$, knowledge $\theta^t_{\ell_{\texttt{NK}}}$ based on the top KCs covered $K_{r_x}$ using their presence $i_{r_x}$, and depth of topic coverage $d_{r_x}$.


\section{\texttt{PEEKC} Dataset}

\begin{figure*}[ht]
\begin{center}
    \centerline{\includegraphics[width=.9\linewidth]{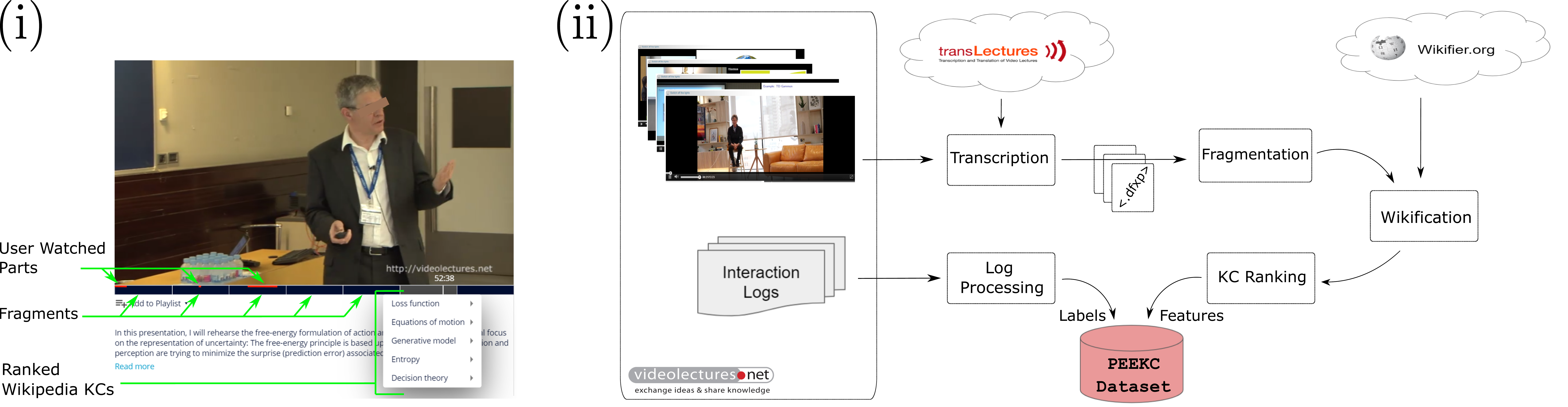}}
    \caption{(i) Visual representation of the data items available in the PEEKC dataset where each video is broken into multiple, non-overlapping 5-minute fragments that are linked with ranked Wikipedia-based KCs and (ii) The flow chart presenting how the video data and the learner interaction logs from VLN repository are processed to create the PEEKC dataset.}
    \label{fig:peek_pipe}
\end{center}
\end{figure*}

In this section, we describe how the \textbf{P}ersonalised \textbf{E}ducational \textbf{E}ngagement linked to \textbf{K}nowledge \textbf{C}omponents (PEEKC) dataset is constructed. \figurename{ \ref{fig:peek_pipe}} (ii) outlines the overall process of creating the PEEKC dataset.
PEEKC uses the data from VideoLectures.Net\footnote{\url{www.videolectures.net}} (VLN), a repository of scientific and educational video lectures. VLN repository records research talks and presentations from numerous academic venues (mainly AI and Computer Science). As the talks are recorded at peer-reviewed research venues, the lectures are reviewed and the material is controlled for the correctness of knowledge. Although most lectures consist of one video, some video lectures are broken into more videos (such as a long tutorial).



\subsection{Fragmenting Video Transcripts}

First, the videos in VLN repository are transcribed to its native language using the \emph{TransLectures} project\footnote{\url{www.translectures.eu}}. Then, the non-English lecture videos are translated into English as we will use English Wikipedia for entity linking.
Once the transcription/translation is complete, we partition the transcript of each video into multiple \emph{fragments} where each fragment covers approximately 5 minutes of lecture time (5000 characters).
Having 5-minute fragments allows us to break the contents of a video into a more granular level while making sure that there is sufficient amounts of information while keeping fragment length at a favourable value in terms of retaining viewer engagement \cite{Guo_vid_prod}.

\subsection{Wikification of Transcripts}
In order to identify the Knowledge Components (KCs) that are contained in different video fragments, we use Wikification \cite{wikifier}. This allows annotating learning materials with humanly interpretable KCs (Wikipedia concepts) at scale with minimum human-expert intervention. This setup will make sure that recommendation strategies built on this dataset will be technologically feasible for web-scale e-learning systems. 


\subsection{Knowledge Component Ranking}

As per \cite{wikifier}, Wikification produces two statistical values per annotated KC, $c$, namely, \emph{PageRank} and \emph{Cosine Similarity} scores. 

\subsubsection{PageRank score} 

is calculated by constructing a semantic graph where semantic relatedness ($SR(c,c')$) between Wikipedia concept pairs $c$ and $c'$ in the graph are calculated using equation \ref{eq:wiki_sr} and running PageRank on this graph.
\begin{equation}\label{eq:wiki_sr}
        SR(c,c') = \frac{\log(max(|L_c|, |L_{c'}|) - \log(|L_c \cap L_{c'}|)}
        {\log |W| - log (min(|L_c|, |L_{c'}|)}
\end{equation}
where $L_{c}$ represents the set of Wiki concepts with inwards links to Wikipedia concept $c$, $|\cdot|$ represents the cardinality of the set and $W$ represents the set of all Wikipedia topics.
PageRank algorithm\cite{pagerank} leads to heavily connected Wikipedia topics (i.e. more semantically related) within the lecture to get a higher score.

\subsubsection{Cosine Similarity score} 
is used as a proxy for topic coverage within the lecture fragment \cite{truelearn}. This score $cos(s_{tr}, c)$ between the \emph{Term Frequency-Inverse Document Frequency (TF-IDF)} representations of the lecture transcript $s_{tr}$ and the Wikipedia page $c$ is calculated based on equation \ref{eq:wiki_cos}:
\begin{equation}\label{eq:wiki_cos}
        cos(s_{tr}, c) = \frac{\texttt{TFIDF}(s_{tr}) \cdot \texttt{TFIDF}(c)}
        {\|\texttt{TFIDF}(s_{tr})\| \times \|\texttt{TFIDF}(c)\|}
\end{equation}
where $\texttt{TFIDF}(s)$ returns the TF-IDF vector of the string $s$ while $||\cdot||$ represents the norm of the TF-IDF vector.

The authors of \cite{wikifier} recommend that a linearly weighted sum between the PageRank and Cosine score can be used for ranking the importance of Wikipedia concepts.


We empirically find weighting 0.8 on PageRank and 0.2 on Cosine similarity is most suitable. The ranked KCs are used to identify the five top-ranked KCs for each lecture fragment. 
\figurename{ \ref{fig:dataset_stats}(ii)} provides a word cloud of the most dominant KCs in the PEEKC dataset. It is evident that the majority of KCs associated with the lecture fragments in this dataset are related to artificial intelligence and machine learning, making this dataset ideal for training personalisation models for AI education.

\subsection{Anonymity}

We restrict the final dataset to lectures with views from at least five unique users to preserve k-anonymity \cite{orcas_dataset}. Also, we report the timestamp of user view events in relation to the earliest event found in the dataset obfuscating the actual timestamp. We report the smallest timestamp in the dataset $t_0$ as 0s and any timestamp $t_i$ after that as $t_i - t_0$. This allows us to publish the real order and the differences in time between events without revealing the actual timestamps. Additionally, the lecture metadata such as title and authors are not published to preserve their anonymity. The motivation here is to avoid video presenters having unanticipated effects on their reputation by associating implicit learner engagement with their content.

\subsection{Labels}
The user interface of VLN website also records the video-watching behaviors of its users (see \figurename{ \ref{fig:peek_pipe} (i)}). 
We create a binary target label based on 
\emph{video watch time} commonly used as a proxy for video engagement in both non-educational \cite{Covington2016,beyondviews} and educational \cite{Guo_vid_prod,truelearn} contexts. Normalised learner watchtime $\overline{e}^t_{\ell,r}$ of learner $\ell$ with video fragment resource $r_i$ at time point $t$ is calculated as per equation \ref{eq:norm_engage}.

\begin{equation} \label{eq:norm_engage}\centering 
\overline{e}^t_{\ell,r_i} = W(\ell,r_i) / D(r_i),
\end{equation}

where $\overline{e}^t_{\ell,r_i} \in \{0,1\}$, $W(\cdot)$ is a function returning the \emph{watch time} of learner $\ell$ for resource $r_i$ and $D(\cdot)$ is a function returning the duration of lecture fragment $r_i$. The final label $e^t_{\ell,r_i}$ is derived by discretising $\overline{e}^t_{\ell,r_i}$ where $e^t_{\ell,r_i} = 1 $ when $\overline{e}^t_{\ell,r_i} \geq .75$ and $e^t_{\ell,r_i} = 0$ otherwise. This rule is motivated by the hypothesis that a learner should watch approximately 4 out of 5 minutes of a video fragment in order to acquire knowledge from it \cite{truelearn}.

\subsection{Final Dataset}

\begin{figure}[ht]
\begin{center}
    \centerline{\includegraphics[width=.7\columnwidth]{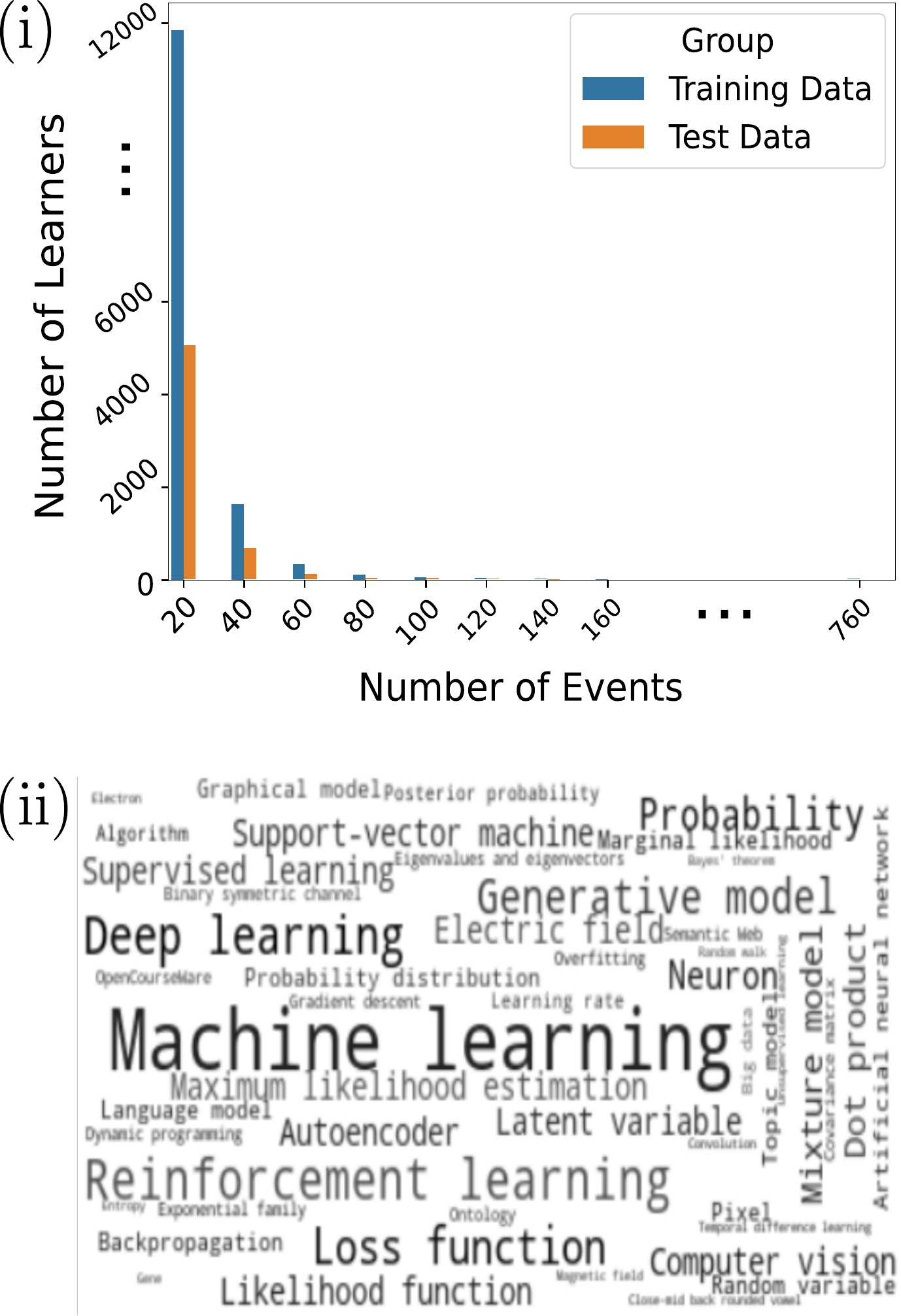}}
    \caption{Characteristics of the PEEKC dataset: (i) number of learners in the training/test dataset based on the number of events in their sessions and (ii) wordcloud depicting the most frequent Wikipedia-based KCs showing the dominance of AI and ML concepts in the dataset.}
    \label{fig:dataset_stats}
\end{center}
\end{figure}

The final PEEKC dataset consists of 290,535 interaction events from 20,019 distinct users with at least five watch events. These learners engage with 8,801 unique lecture videos partitioned into 36,408 fragments (4.14 fragments per video). The learner population in the dataset is divided into \emph{Training} (14,050 learners) and \emph{Test} (5,969 learners) datasets based on a 70:30 split. The label distribution in the dataset is also relatively balanced with only 56.35\% of the labels being positive. As shown in \figurename{ \ref{fig:dataset_stats} (i)}, the majority of learners in the dataset have a relatively small number of events (under 80) making this dataset an excellent test bed for personalisation models designed to work in data-scarce environments. VLN repository mainly publishes videos relating to AI and Machine Learning leading to a learner audience who visit to learn about these subjects. This fact is confirmed by \figurename{ \ref{fig:dataset_stats}} where it shows that the dataset is dominated by events with AI and ML-related KCs. The dataset is available publicly\footnote{\url{https://github.com/sahanbull/PEEKC-Dataset}}.


    

\begin{table}[] \small
\caption{Different columns included in the PEEKC Dataset.}
\label{tab:features}
\centering 
\begin{tabular}{clc}
\hline
Column & Description & Data Type\\
\hline
1 & Video Lecture ID & Integer\\
2 & Video ID & Integer \\
3 & Part ID & Integer\\
4 & Timestamp & Integer \\
5 & User ID & Integer \\
6,8,10,12,14 & Knowledge Component IDs & Integer\\
7,9,11,13,15 & Topic Coverage & Floating Point\\ 
16 & Label & Binary\\
\hline
\end{tabular}
\end{table}

\section{\texttt{truelearn} Library}

This section describes the architecture of the TrueLearn Python library. While TrueLearn provides a probability that can be mapped to a binary outcome (engaging/not engaging), the probability prediction on different videos ranks them, creating personalised recommendations.


\subsection{Architecture}

The TrueLearn library consists of six modules.



\subsubsection{Datasets}
The dataset module integrates tools for downloading and parsing learner engagement datasets. Currently, the PEEKC dataset is integrated. 

\subsubsection{Pre-processing}
The pre-processing module contains utility classes for extracting content representations from educational materials. The extracted representations become KCs that can be used with IRT, KT and TrueLearn models. At present, utility functions for Wikification are included. 

\subsubsection{Models}
This module houses the class that can store the learner model. In this context, the learner model refers to the data structure storing the learner state (e.g. knowledge/ interest). This learner model is loosely coupled with the learning algorithms which makes this object reusable with other learning algorithms that go beyond the TrueLearn algorithms.

\subsubsection{Learning}
This module contains machine learning algorithms that can perform training and prediction of learner engagement with transcribed videos. 
For training,  \texttt{fit} function is used. For prediction, \texttt{predict} and \texttt{predict\_proba} functions are used.
Currently, a set of baselines and the TrueLearn algorithms \cite{bulathwela2022sus} are included. 

\subsubsection{Metrics}
Classification metrics accuracy, precision, recall and F1-score are built as the task is posed as a classification task in prior work \cite{truelearn}. The module is easily extendable to regression and ranking metrics. 

\subsubsection{Visualisations}
To effectively present the learner state, \emph{nine} different visualisations shown promising in prior work on user interaction \cite{mti6060042} have been developed. 
\figurename{ \ref{fig:bubble}} provides a preview of one of the common visualisations. Seven (out of nine) interactive visualisations allow the learner to click and hover over the output to explore more details.

\subsection{Visualising the Learner State}

Our approach was guided by a thorough examination of seminal research on impactful learning visualisations. Among others, the interactive visualisations designed in our Python library takes into account the goals of self-actualisation as detailed in the EDUSS framework \cite{mti6060042}. 
Additionally, the visualisations utilise user-friendly cues and conventions (colours/intensity of colour, shape size etc.) to minimise the cognitive load. 
Based on user preferences found on learning visualisations \cite{visualisationscomparison}, the i) bar plot, ii) dot plot, iii) pie plot, vi) tree plot v) radar plot, vi) rose plot vii) bubble plot viii) word plot and ix) line plot were chosen to be implemented. Figure \ref{fig:bubble} previews the learner state of one of the learners. 

TrueLearn algorithms model learner skill states as Gaussian variables with mean (state estimate) and variance (estimate uncertainty). The bar plot and dot plot use the bar/dot for skill mean while mapping the uncertainty as a confidence interval. The radar plot uses the radius of the radar as the skill estimate. The pie plot and tree plot use the area of the pie to represent skill mean while using the intensity of the colour (dark to light) for uncertainty. Rose plot, in addition to using the radius and colour intensity for mean and uncertainty respectively, uses the area of the pie to depict the number of video fragments that contributed to the skill estimate. The bubble plot and word plot use the size of the skill shape to represent the mean estimate, while the bubble plot uses the colour intensity to depict model uncertainty (\figurename{ \ref{fig:bubble}}). The line plot uses the x-axis as the time to show how a skill evolved over time. Popular learner models such as KT and IRT do not provide uncertainty for the skill estimate. Still, the implemented visualisations work without skill uncertainty values. 

\section{Experiments and Results}

We use the experimental protocol used in \cite{bulathwela2022sus} for our experiment. 

\subsubsection{Baselines} We use a wide range of baselines that are i) exclusively content-based and ii) maintain a concept-based user model \cite{zarrinkalam2020extracting}. As content-based models, we use i) KC cosine similarity-based (\texttt{Cosine}), Jaccard similarity based on ii) KC intersection ($\texttt{Jaccard}_{\mathcal{C}}$) and iii) user intersection ($\texttt{Jaccard}_{\mathcal{U}}$). As concept-based user models, we use iv) TF (Binary), which counts the number of times a concept was encountered and v) TF (Cosine), which aggregates the cosine scores for skills in PEEKC dataset over time and vi) online Knowledge Tracing model (KT) \cite{bishopsnewbook}. 

\subsubsection{TrueLearn Models} We use the three TrueLearn models implemented in the library, namely, i) TrueLearn Interest, capturing interests, ii) TrueLearn Novelty, capturing knowledge and novelty and iii) TrueLearn INK, combining interests, knowledge and novelty.


\subsubsection{Data and Evaluation} For each learner, its engagement at time $t$ is predicted using its events at times $1$ to $t-1$. We used the hold-out validation (70\% train/ 30\% test) technique in our experiment where the training data in PEEKC is used for hyperparameter tuning of models. The best hyperparameter combination based on the F1-Score is identified and used with the test set to evaluate the reported performance. Since the engagement is labelled as a binary label in the PEEKC dataset, accuracy, precision, recall, and F1 score are reported. 

\subsection{Empirical Evaluation}

The results are reported in Table \ref{tab:results}. Our experiments i) guarantee the correctness of the library implementation and ii) demonstrate the predictive capabilities of the web-scale online learning models with comparable baselines. 




\begin{table}[] \small
\caption{Performance of TrueLearn algorithms are evaluated using Precision (Prec.), Recall (Rec.) and F1 Score (F1). The best and second best performance is indicated in \textbf{bold} and \emph{italic} faces respectively. TrueLearn models outperform the best baseline most times ($p< 0.01$ in a one-tailed paired t-test are marked with $\cdot^{*}$).}
\label{tab:results}
\begin{tabular}{cccccc}
\hline
& Model              & Acc. & Prec. & Rec. & F1    \\
\hline
&\texttt{Cosine}& {55.08} & {57.86} & 58.45 & 54.06 \\
&$\texttt{Jaccard}_{\mathcal{C}}$& 55.46 & 57.81 & 60.36 & 55.03 \\
Baselines&$\texttt{Jaccard}_{\mathcal{U}}$ & 64.06 & 57.85 & {72.76} & {61.22} \\
&TF (Binary)	& 55.19 &	56.71 & 66.60 &	57.38 \\
&TF (Cosine) & 55.11 &	56.75	& 65.95	 & 57.11 \\ 
&KT & 54.99 & 53.25 & 28.56 & 34.51 \\
\hline
&Interest & 58.13    & 52.08     & \textit{78.61}$^{*}$  & 63.00$^{*}$ \\ 
TrueLearn & Novelty  & \textit{64.78}$^{*}$	  & \textit{58.52}$^{*}$  & \textbf{80.91}$^{*}$  & \textbf{65.53}$^{*}$ \\
&INK  & \textbf{78.32}$^{*}$	  & \textbf{64.32}$^{*}$	  & {64.03}  & \textit{64.00}$^{*}$ \\
\hline
\end{tabular}
\end{table}


\begin{figure}[]
\begin{center}
\centerline{\includegraphics[width=.7\linewidth]{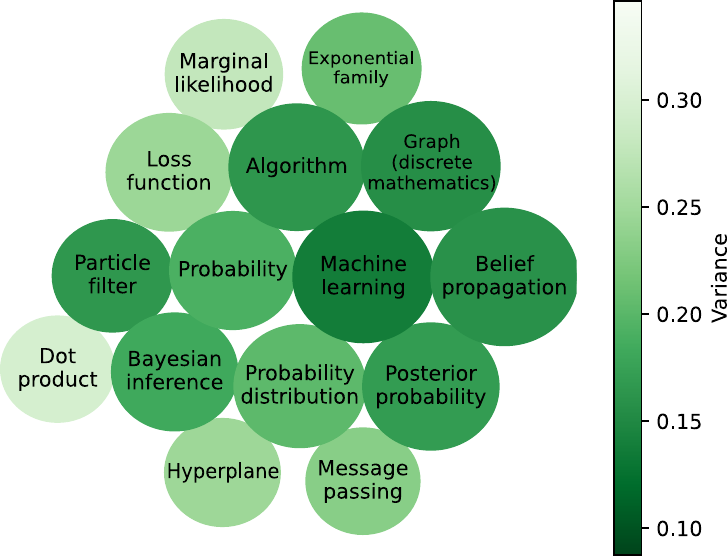}}
 \caption{The 15 most knowledge acquired KCs of a learner in a bubble plot. The size of the circle aligns with the KC mean and the intensity of the colour maps to the variance.}
\label{fig:bubble}
\end{center}

\end{figure}


\section{Discussion}

The contributions in this work span over a novel dataset, an OLM library and empirical experiments. 

\subsection{PEEKC for AI Education}
As per the wordcloud in \figurename{ \ref{fig:dataset_stats} (ii)}, the videos in the PEEKC dataset are swamped with AI and ML-related KCs. PEEKC's data source, VLN repository extensively visits AI conferences which causes this effect. But, this makes PEEKC a perfect dataset to understand in-the-wild video-based knowledge acquisition, making this dataset an excellent resource for training personalisation models for AI education as we have done in table \ref{tab:results}. While we haven't demonstrated here, the dataset is potentially valuable for unsupervised tasks such as pre-requisite identification and hierarchical structuring of concepts in AI. A key benefit of PEEKC dataset is the ability to use the methodology in \figurename{ \ref{fig:peek_pipe}} with VLN repository to create larger datasets in the future.

\subsection{TrueLearn Performance and Visualisations}

Table \ref{tab:results} clearly shows the superiority of the TrueLearn model implementations in comparison to the comparative baselines aligning with prior work's evidence \cite{bulathwela2022sus}. The most expensive, \verb|fit| function measures $0.005\pm0.0036s$ to execute in an Apple M1 3.2GHz CPU. TrueLearn states of each learner in the dataset can be constructed in parallel because there are no inter-learner dependencies, also making the model privacy-preserving by design. The performance in table \ref{tab:results} coupled with event counts in \figurename{ \ref{fig:dataset_stats}} shows the data efficiency of TrueLearn, which is able to learn from a very small number of events.  

The library is designed to enable learners to effortlessly generate dynamic and static visualisations, promoting a learner-centric, self-regulated study experience \cite{bull2008metacognition}. \figurename{ \ref{fig:bubble}} previews the learner knowledge state from one of the learners in the PEEKC test data. It is seen that the visualisation demonstrates how the user is acquiring knowledge in AI-related topics, specifically, in the area of Bayesian modelling. 
The EDUSS framework-inspired visualisations support development by translating skill-level predictions into visual progress indicators encouraging continuous learning. Understanding is enhanced by providing insights into learners' interaction patterns, promoting self-awareness \cite{mti6060042}. The transparency of the user model allows learners to scrutinise their learning progress and engage critically with the underlying data (while ways to provide this feedback to the model is still an open research area). Finally, by making the visualizations shareable, the library fosters social interaction, adding a community dimension to the learning experience and improving its usability to other stakeholders of education (e.g. parents and teachers).

\subsection{Library Design, Stability and Maintainability}

The adherence to prior work-inspired good practices and API design makes using the TrueLearn library developer-friendly for both offline experimentation and e-learning system integration. Using a collection of base classes that define a common interface and shared functionality similar to scikit-learn-like estimators with \verb|fit| and \verb|predict| functions \cite{buitinck2013api} reduces the learning curve while the design allows elaborate type and value checks when the hyperparameters of the classifiers are modified, ensuring the robustness of the classifier implementation.
The decoupling of the feature-engineering from modelling modules allows users to extend the capability of the library with new KC extraction functions (beyond Wikification), new online learner models and open learner visualisations without having to worry about interdependencies. Detailed instructions are provided to developers with style guides and design principals \footnote{Contributing: \url{https://truelearn.readthedocs.io/en/latest/dev/}} to ease contribution. The core research team is committed to maintaining the library in the future while providing further guidance and code reviews when extending the capabilities of the library.   
Furthermore, TrueLearn benefits from 100\% test coverage achieved through a combination of integration, unit, and documentation tests. 


Code consistency and readability are further enhanced by following the PEP 8 guidelines \cite{pep8}, which define a set of best practices for Python code. Extensive documentation
of the modules and classes with in-context code examples describes relevant information for both a potential developer and a contributor to familiarise themselves with the library. Developers have already started adapting this library to learning platforms \cite{x5learn}. We aim to objectively assess their experience by surveying them \cite{piccioni2013empirical,nadi2023selecting}.  


\subsection{Relevance, Impact and Limitations}

{Modelling learner state in a humanly intuitive manner, requiring minimal data and exclusively relying on individual user actions, TrueLearn offers a transparent learner model that respects the privacy of its users and can scale to lifelong education. The development of the TrueLearn library aims to provide both the research and developer communities with the opportunity to seamlessly use the TrueLearn family of models in their work. The learner models utilise Wikipedia-based entity linking to create KCs based on a publicly available knowledge base. The content annotation can also scale to thousands of materials created in different modalities (video, text, audio etc.).}

{The impact of TrueLearn is two-fold. For developers and researchers, the TrueLearn library employs a design that conforms with popular machine learning libraries. The documentation is extensive and contains detailed examples that help the implementation. For developers and educators, probabilistic graphical models that are data efficient and humanly intuitive are available to be used in their downstream systems. The engagement predictions (between values 0 and 1) can be used to rank videos for personalised learning. Combined with the PEEKC dataset, hyperparameters for a new system can be trained beforehand and deployed in a new online learning platform. The online learning algorithm updates the learner state in real-time helping better personalisation. A platform implementing TrueLearn can scale to a large population of users and support them through lifelong education due to the large number of KCs it can support.}



{While getting inspired by scikit-learn library, the learning algorithms in TrueLearn library are not compatible with some helper functions available in scikit-learn (such as grid search) and pandas libraries at this point. Building seamless compatibility with these utilities will enable the TrueLearn library to be adopted by a wider audience while minimising the development effort required to support such powerful features. While the visualisations implemented are time-tested \cite{visualisationscomparison,mti6060042}, their success with TrueLearn representations has room for rigorous understanding via user studies. The exclusive support of online learning algorithms can also be seen as a limitation of the current library as many batch learning algorithms are proposed for educational recommendation and engagement modelling \cite{intervention_bkt,ensemble_kt}. While learner engagement is a prerequisite for learning, it is noteworthy that learner engagement doesn't imply learning. The library also does not support state-of-the-art deep learning algorithms \cite{deep_kt,pardos_serendipity} that may be useful where interpretability and learner state visualisation are not the top priority.}



\section{Conclusion}

{This work presents \textit{PEEKC} dataset and \textit{TrueLearn} Python library, creating a valuable toolbox for engagement modelling with AI-related educational videos. The library contains several online learning models, which model multiple factors influencing learner engagement. It also packages a set of visualisations that can be used to interpret the learner's interest/knowledge state. The learner representations and state visualisations are comparable to outputs of knowledge tracing models, except TrueLearn uses watch time interactions rather than relying on test taking. The empirical results demonstrate that the implementation of the library achieves similar performance to the prior work. The new implementation encourages educational data mining practitioners to use this library to incorporate educational video recommendations in e-learning systems. Researchers are encouraged to extend this library with new datasets and online learning algorithms for learner engagement modelling.}

The immediate future work entails improving learner state visualisations via user studies. 
Integrating the library into a real-world e-learning platform \cite{10.1145/3397482.3450721} is a top priority. 
Extending the current framework to podcasts and other information content while incorporating other feedback forms like educational questions \cite{bulathwela2023scalable,bulath2023neurips} remains in the future roadmap. 
In the long term, we aim to add more general informational recommendation algorithms to the library and mobilise the research community to contribute various models, pre-processing techniques and evaluation metrics that the library can benefit from. 

\section{Acknowledgments}
This work is partially supported by the European Commission-funded project "Humane AI: Toward AI Systems That Augment and Empower Humans by Understanding Us, our Society and the World Around Us" (grant 820437) and the X5GON project funded from the EU's Horizon 2020 research programme grant No 761758.

\bibliography{aaai24}

\newpage
\section{Supplementary Material}

\subsection{Detailed Problem Setting}

A learner $\ell$ in learner population $L$ interacting with a series of educational resources $S_\ell \subset \{r_1, \ldots, r_{R}\}$ where $r_x$ are \emph{fragments/parts} of an educaitonal video $v$. The watch interactions happen over a period of $T$ time steps, $R$ being the total number of resources in the system.
In this system with a total $N$ unique knowledge components (KCs), resource $r_x$ is characterised by a set of top KCs or topics $K_{r_x} \subset \{1, \ldots, N \}$. We assume the presence $i_{r_x}$ of KC in resource $r_x$ and the degree $d_{r_x}$ of KC coverage in the resource is observable.

The key idea is to model the probability of engagement $e_{\ell, r_x}^{t} \in \{ 1, -1\}$ between learner $\ell$ and resource $r_x$ at time $t$ as a function of the learner interest $\theta^t_{\ell_{\texttt{I}}}$, knowledge $\theta^t_{\ell_{\texttt{NK}}}$ based on the top KCs covered $K_{r_x}$ using their presence $i_{r_x}$, and depth of topic coverage $d_{r_x}$.

\begin{figure}[H]
\begin{center}
\centerline{\includegraphics[width=0.65\columnwidth]{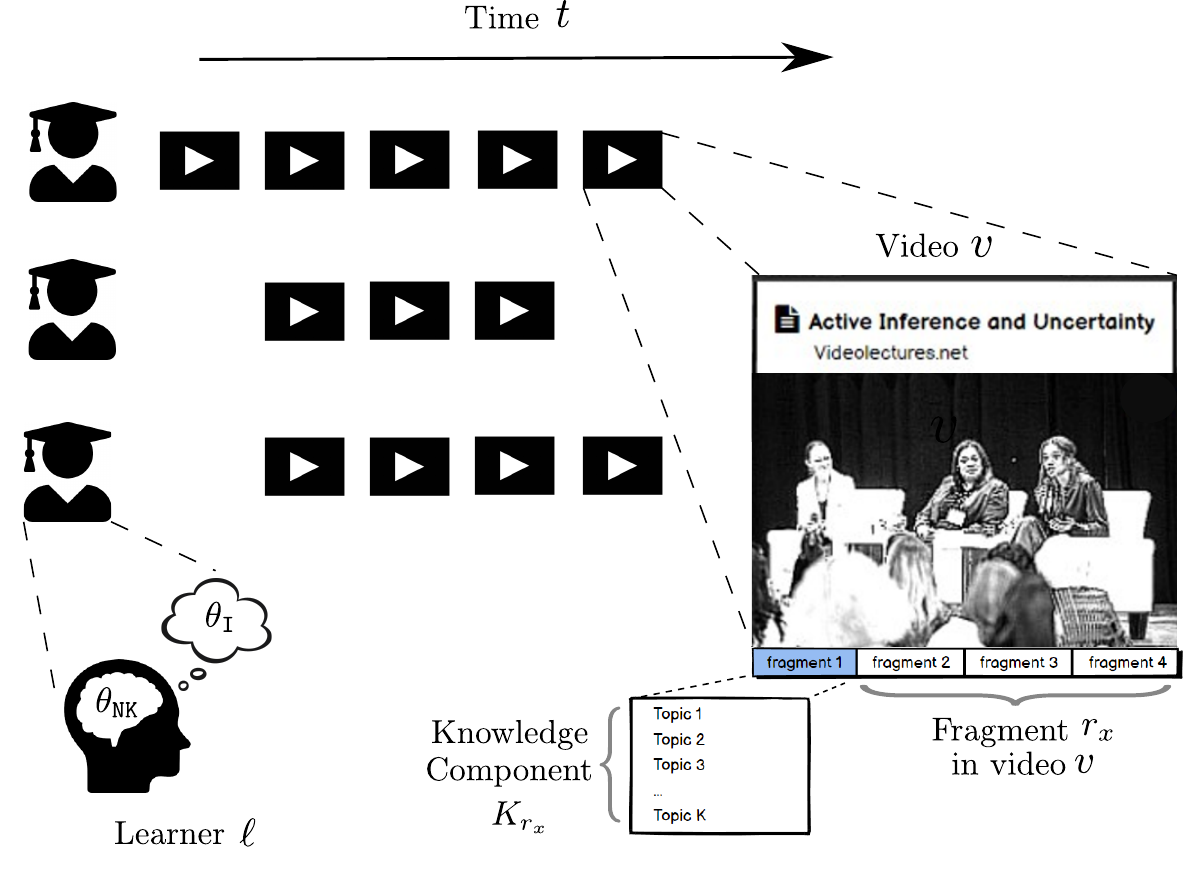}}
\caption{Visual illustration of the problem setting where learner $\ell$, with knowledge (that allows them to tackle novel content) $\theta_\texttt{NK}$ and interests $\theta_\texttt{I}$ is watching fragments of educational videos $r_x$ containing different knowledge components $K_{r_x}$ over time $t$.}
\label{fig:semantic_rel_problem}
\end{center}
\end{figure}

\subsection{Baseline Models}

PEEKC is the first of its kind, a dataset that records in-the-wild  engagement of informal learners with video lecture fragments. Due to the novelty of this dataset, we struggle to find already published baselines, except for the TrueLearn family of algorithms. 
For the sake of comparing its predictive performance,  we also propose a set of baselines that are based on content-based and collaborative filtering.

\paragraph{Content-based Similarity} Content-based filtering can measure the similarity between two items. We compute a similarity value, $sim(r^{t-1}_{\ell,r_i},r^{t}_{\ell,r_i})$ between two consecutive lecture fragments $r^{t-1}_{\ell,r_i}$ and $r^{t}_{\ell,r_i}$ in the learner  $\ell$'s session. We use this similarity value to make an engagement prediction $\hat{e}^t_{\ell,r_i}$ based on equation \ref{eq:cbf_baseline}.

\begin{equation} \label{eq:cbf_baseline}\centering 
    \hat{e}^t_{\ell,r_i} = 
    \left\{ 
      \begin{array}{ c l }
        1                 &  if\ sim(r^{t-1}_{\ell,r_i},r^{t}_{\ell,r_i}) \geq threshold \\
        0                 & otherwise
      \end{array}
    \right.
\end{equation}

In this case, we investigate two similarity measures, namely 1) \emph{Cosine}, 2) \emph{Concept-based Jaccard} and \emph{User-based Jaccard}. When computing cosine similarity, we represent each video fragment using the bag of concepts representation where the concepts are the superset of Wikipedia concepts mentioned in the dataset. The values in this sparse vector are the cosine similarities between the respective Wikipedia concept and the lecture fragment transcript as per equation \ref{eq:wiki_cos}.

\begin{equation}\label{eq:wiki_cos}
        cos(s_{tr}, c) = \frac{\texttt{TFIDF}(s_{tr}) \cdot \texttt{TFIDF}(c)}
        {\|\texttt{TFIDF}(s_{tr})\| \times \|\texttt{TFIDF}(c)\|}
\end{equation}

An alternative approach to finding concept-wise similarity is Jaccard similarity. Concept-based Jaccard similarity $\texttt{Jaccard}_{\mathcal{C}}(r^{t-1}_{\ell,r_i}, r^{t}_{\ell,r_i})$,  between lecture fragments $r^{t-1}_{\ell,r_i}$ and $r^{t}_{\ell,r_i}$ is computed based on equation \ref{eq:concept_jaccard}.

\begin{equation} \label{eq:concept_jaccard}\centering 
    \texttt{Jaccard}_{\mathcal{C}}(r^{t-1}_{\ell,r_i}, r^{t}_{\ell,r_i}) = 
    \frac{\mathcal{C}(r^{t-1}_{\ell,r_i}) \cap \mathcal{C}(r^{t}_{\ell,r_i})}
    {\mathcal{C}(r^{t-1}_{\ell,r_i}) \cup \mathcal{C}(r^{t}_{\ell,r_i})}
\end{equation}  
where $\mathcal{C}(\cdot)$ is a function that returns the set of Wikipedia concepts in resource $r_i$

Similarly, one can also measure the similarity between two lecture fragments based on how many learners interact with both lecture fragments. The user interactions in the training dataset is used exclusively to learn the similarity matrix in order to avoid data leakage. In this approach, we can calculate the user-wise Jaccard similarity $\texttt{Jaccard}_\mathcal{U}(r^{t-1}_{\ell,r_i}, r^{t}_{\ell,r_i})$, as per equation \ref{eq:user_jaccard}. 

\begin{equation} \label{eq:user_jaccard}\centering 
    \texttt{Jaccard}_\mathcal{U}(r^{t-1}_{\ell,r_i}, r^{t}_{\ell,r_i}) = 
    \frac{\mathcal{U}(r^{t-1}_{\ell,r_i}) \cap \mathcal{U}(r^{t}_{\ell,r_i})}
    {\mathcal{U}(r^{t-1}_{\ell,r_i}) \cup \mathcal{U}(r^{t}_{\ell,r_i})}
\end{equation}  
where $\mathcal{U}(\cdot)$ is a function that returns the set of learners that interacted with resource $r_i$

\paragraph{Knowledge Tracing (KT)}  

KT builds a learner representation of the knowledge of the learner. This learning model is then used in predicting the engagement of learner $\ell$ with lecture fragment resource $r_i$ at time $t$. As the PEEKC dataset has a temporal dimension, we reformulate the KT algorithm into an online learning graphical model inspired by the reformulation found in prior work.  The skill variables in the KT model are Bernoulli variables ($\theta^t_{\ell,c} \sim \texttt{Bernoulli}(\pi^t_{\ell,c})$), assuming that a learner $\ell$ would have either mastered a skill/ concept $c$ or not (represented by probability $\pi^t_{\ell,c}$). Skills are initialised ($\theta^0_{\ell,c}$) using a $\texttt{Bernoulli}(.0)$ prior, assuming that the latent skill is not mastered in the beginning. A noise factor similar to what is found in the conventional KT model is added to this model and is tuned using a grid search.

\newpage
\subsection{Number of Knowledge Components Published with the PEEKC Dataset}

\begin{figure}[h]
\begin{center}
    \centerline{\includegraphics[width=1.1\linewidth]{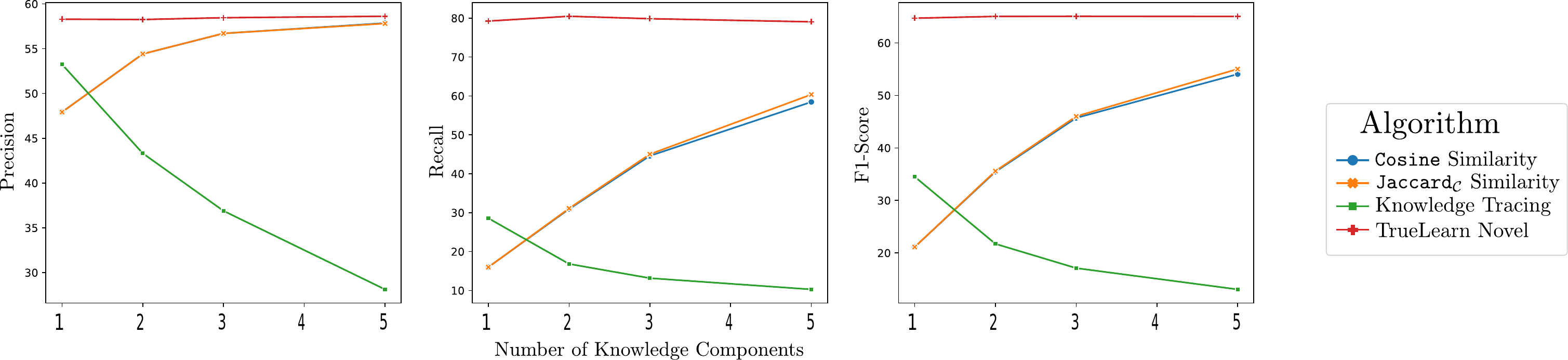}}
    \caption{Predictive performance on PEEKC dataset test data in terms of Precision (left), Recall (middle) and F1-Score (right) for the benchmark models when varying numbers of Knowledge Components (KCs) are used as the content representation. Higher number of topics did not increase the performance of the Cosine and Jaccard models significantly to reach TrueLearn Novelty model}
\end{center}
\end{figure}

\subsection{Hyperparameters for Reproducing Experiments}

\begin{table}[h] \footnotesize
\caption{The hyperparameter combinations that produced the results that are reported in the results section.} 
\centering 
\begin{tabular}{c c l c}
  \hline
    Sec. &  Model & Relevant Hyperparameter & Value \\
  \hline
  1 & & Draw Probability & 0.52\\
  & TrueLearn& Initial Variance & 300.00 \\
  & Interest& Beta & 8.83 \\
  & & Tau & 0.0\\
  & & Best Hyperparameter Set & Max. F1-Score \\
  \hline
 2 & & Draw Probability & 0.52\\
  & TrueLearn& Initial Variance & 0.25 \\
  & Novelty& Beta & 0.42 \\
  & & Tau & 0.0\\
  & & Best Hyperparameter Set & Max. F1-Score \\
  \hline
  3 & & Greedy & True\\
  & & Tau & 0.5\\
  & TrueLearn & Interest Model & Model in Sec. 1 \\
  & INK & Novelty Model & Model in Sec. 2 \\
  & & Best Hyperparameter Set & Max. F1-Score \\
   \hline
\end{tabular}
\end{table}

\subsection{Computational Complexity}

The execution time benchmarks were carried out using a laptop computer with an Apple M1 3.2GHz CPU and 16GB RAM.

\begin{table}[h!] \footnotesize
\caption{The mean duration with the standard error (1.96 $\times$ standard deviation) taken for a function execution of a single event of a single learner in the PEEKC dataset.} 
\centering 
\begin{tabular}{l c c}
  \hline
  Function & TrueLearn Model & Execution Time (s) \\
  \hline
  & Interest & 0.002016 $\pm$ 0.00478
  \\
  \texttt{fit()} & Novelty & 0.002232 $\pm$ 0.00468 \\
   & INK & 0.005000 $\pm$ 0.00712 \\
   \hline
  & Interest & 0.000084 $\pm$ 0.000020   \\
  \texttt{predict\_proba()} & Novelty & 0.000189 $\pm$ 0.000018 \\
   & INK & 0.000330 $\pm$ 0.000025 \\
   \hline
    & Interest & 0.000093 $\pm$  0.000014 \\
  \texttt{predict()} & Novelty & 0.000203 $\pm$  0.000027 \\
   & INK & 0.000335 $\pm$  0.000029  \\
   \hline
\end{tabular}
\end{table}

\newpage
\section{Data Efficiency of TrueLearn Models}


\subsubsection{Improvement of Prediction in the Entire Test Set}

\begin{figure}[h]
\begin{center}
    \centerline{\includegraphics[width=\linewidth]{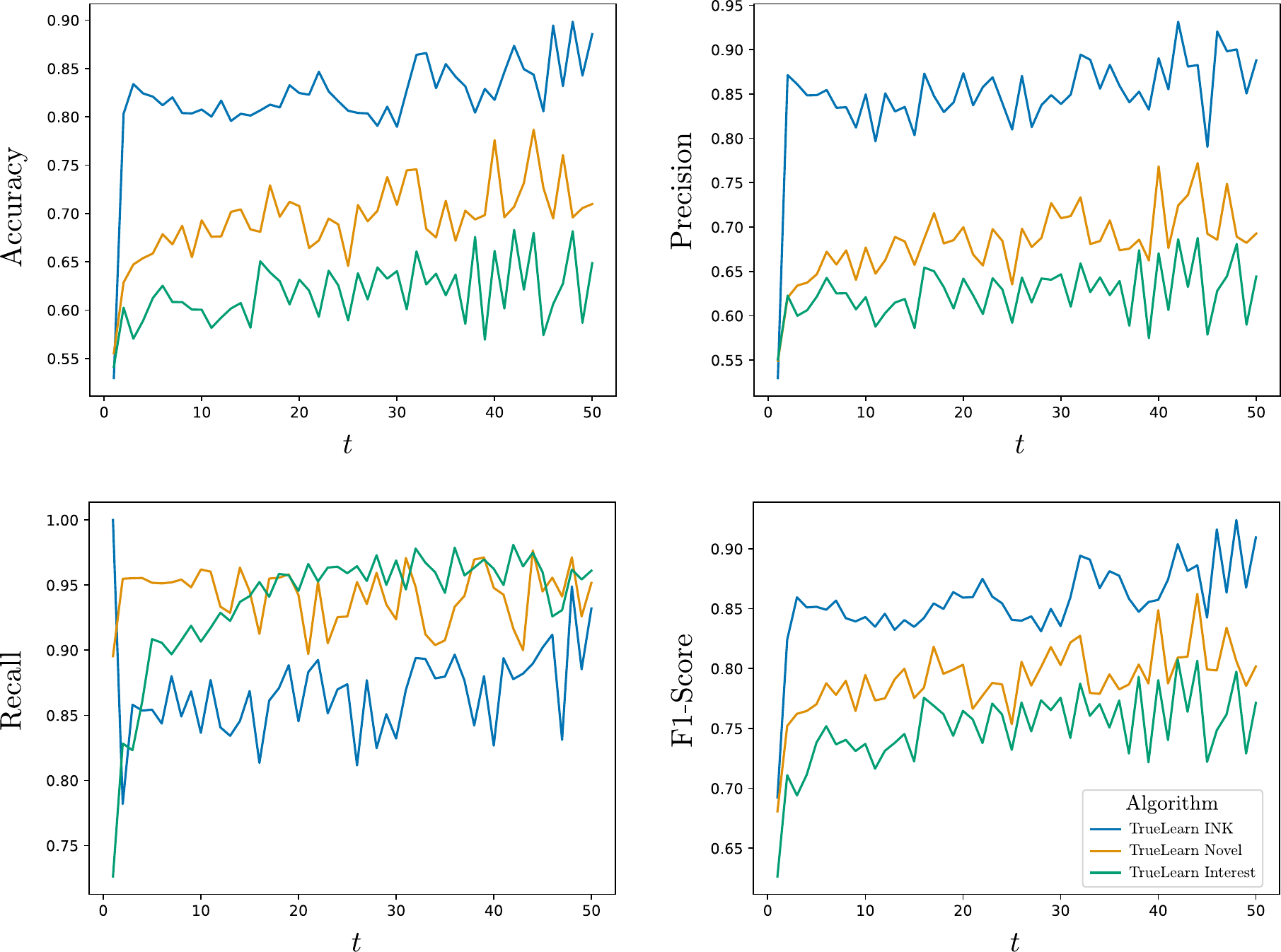}}
    \caption{How the Mean Accuracy, Precision, Recall and F1-Score at time $t$ across all users in the PEECK test set change on TrueLearn Interest (Green), TrueLearn Novel (Yellow) and TrueLearn INK (Blue) Models across the entire dataset.}
     \label{fig:INK_ACC}
\end{center}
\end{figure}

Presented in figure \ref{fig:INK_ACC}

\subsubsection{Improvement of Prediction in the First 10 Events}

\begin{figure}[h] 
\begin{center}
    \centerline{\includegraphics[width=\linewidth]{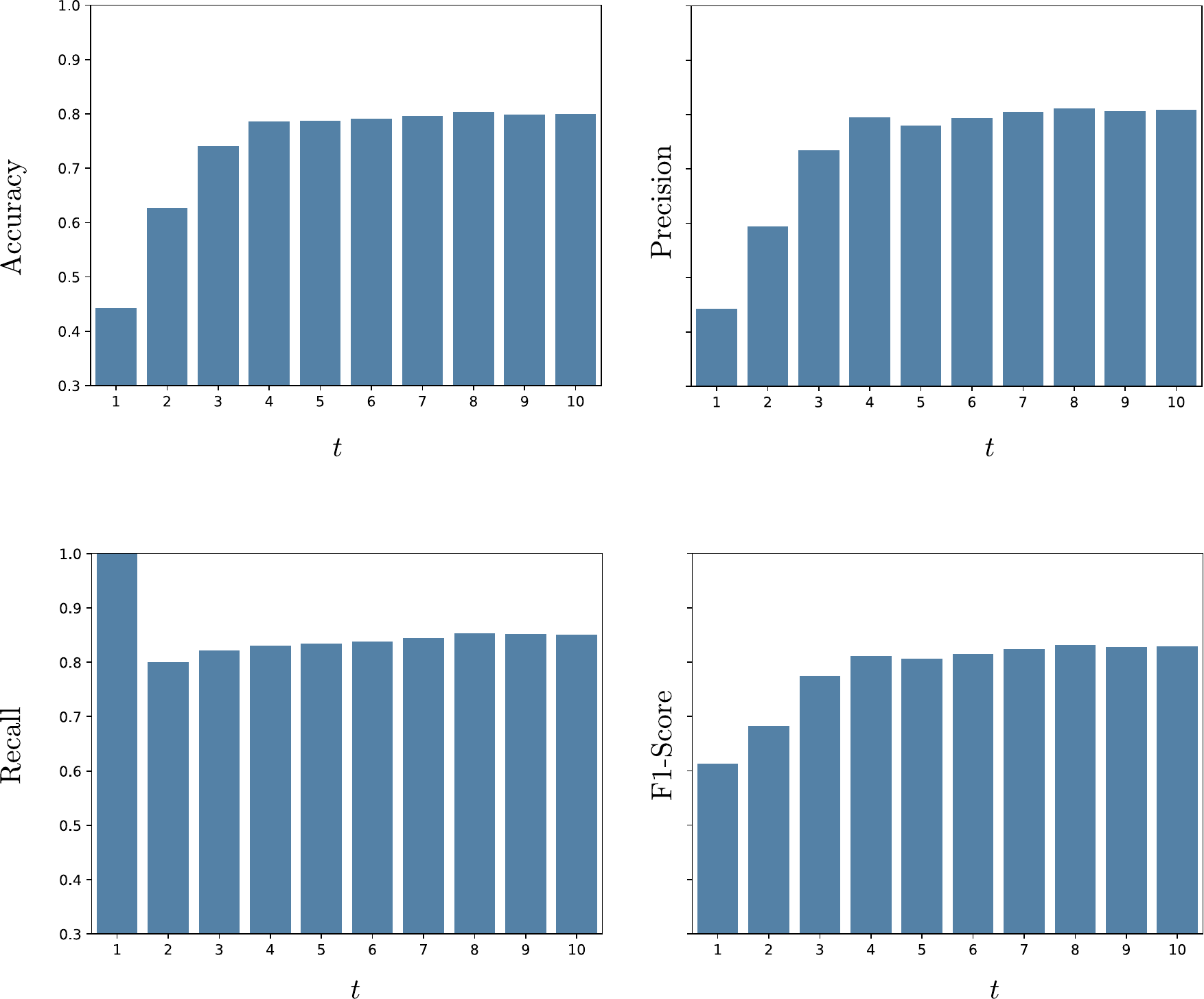}}
    \caption{How the mean Accuracy, Precision, Recall and F1-Score at time $t$ across all the users in the PEEKC test set change the TrueLearn INK Model within the first 10 events.}
    \label{fig:INK_MET}
\end{center}
\end{figure}

Presented in figure \ref{fig:INK_MET}

\newpage

\subsubsection{Subset of Learner State Visualisations}
\begin{figure}[h!]
\begin{center}
    \centerline{\includegraphics[width=\linewidth]{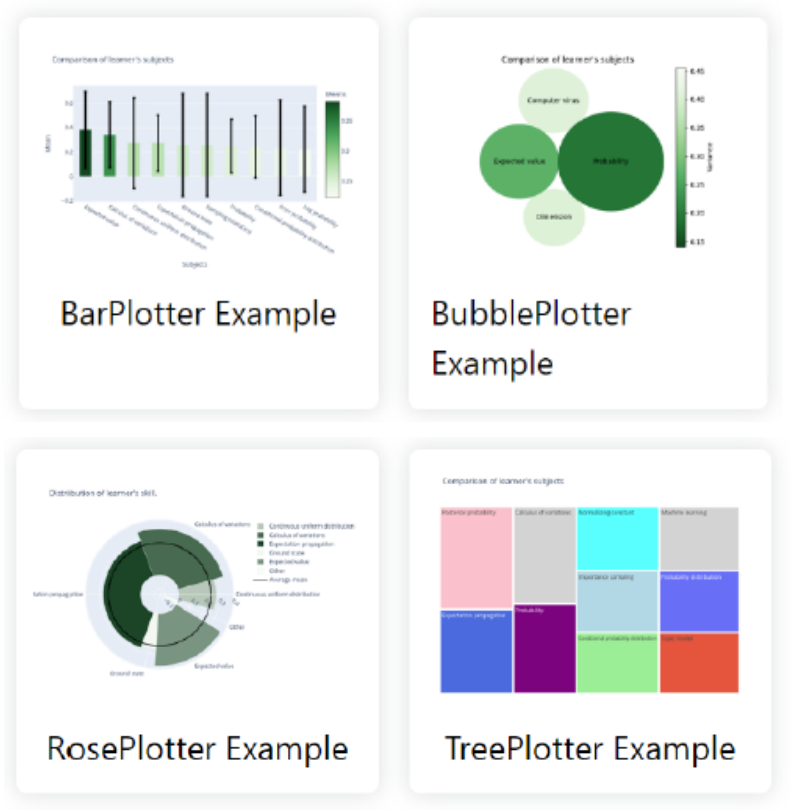}}
    \caption{A subset of the multiple visualisations available in the TrueLearn library to present the learner state in a humanly intuitive way.}
\end{center}
\end{figure}

\newpage
\subsection{TrueLearn Models}

\begin{figure*}[h]
\begin{center}
\centerline{\includegraphics[width=.75\linewidth]{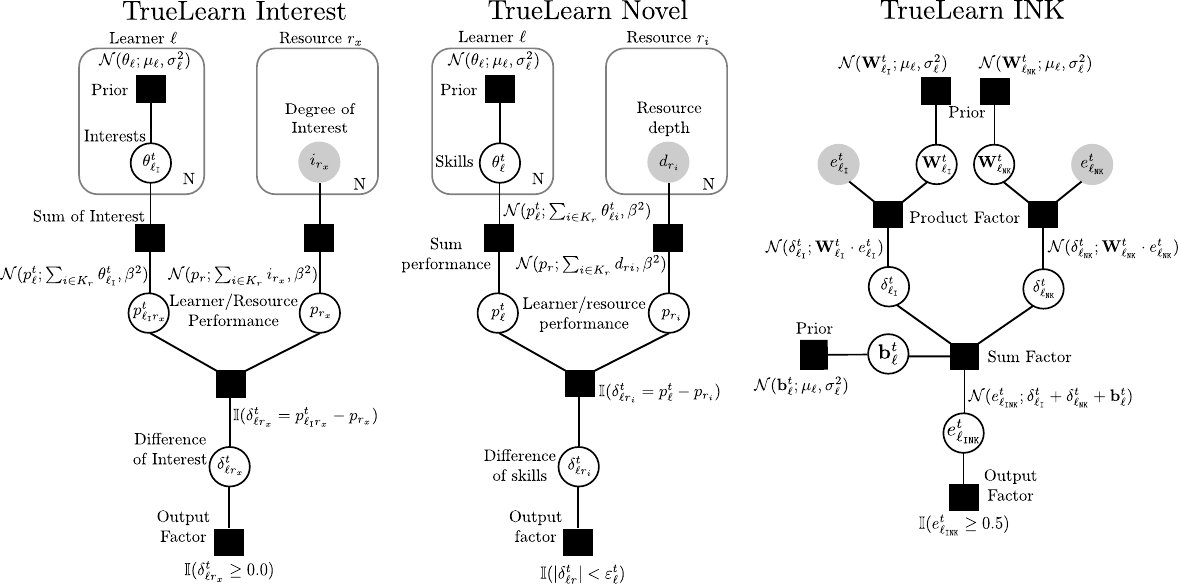}}
\caption{Factor graphs representing the probabilistic graphical models for TrueLearn Interest (left), TrueLearn Novelty (middle) and TrueLearn INK (right) models.}
\end{center}
\end{figure*}

\newpage
\subsection{Detailed Descriptions of the Columns in the PEEKC Dataset}

Detailed descriptions of the columns in the PEEKC dataset are found in table \ref{tab:features}
\begin{table*}[h] \small
\caption{Detailed descriptions of the different columns of the \texttt{train.csv} and \texttt{test.csv} files included in the PEEKC Dataset.}
\label{tab:features}
\centering 
\begin{tabular}{ccl}
\hline
Column Number & Description & Details \\
\hline
1 & Video Lecture ID & An integer ID associated with an individual video lecture \\
2 & Video ID & An integer ID associated to every video belonging to the\\
& & same Video lecture ID (e.g. $1\dots v$ if the lecture has $v$ videos) \\
3 & Part ID & An integer ID associated with each video fragment \\
& & (e.g $1 \dots f$ for a video with $f$ fragments) \\
4 & Timestamp & Timestamp (to the nearest second) when the play event was \\
& &  initiated. \\
5 & user ID & An integer ID associated with each unique learner in the \\
& & dataset PEEKC dataset (IDs in \texttt{train.csv} and \texttt{test.csv} \\ 
& & files are mutually exclusive). \\
6,8,10,12,14 & KC IDs & An integer ID associated with each unique Knowledge \\
& & Component. This ID can be linked to the human-readable   \\
& & Wikipedia concept names \\
7,9,11,13,15 & Topic Coverage & Proxy for coverage of the relevant KC in the fragment of  \\ 
& & interest. KC coverage is the cosine similarity. \\
16 & Label & The binary label $e^t_{\ell,r_i}$, 1 if the learner watched $\geq$ .75 \\
& & of the video  fragment, 0 otherwise.\\
\hline
\end{tabular}
\end{table*}

\end{document}